\documentclass[12pt,a4paper]{article}
\usepackage{amssymb,amsmath}
\usepackage{graphicx,subfigure}
\def\beq{\begin{equation}}
\def\eeq{\end{equation}}
\def\beqr{\begin{eqnarray}}
\def\eeqr{\end{eqnarray}}
\def\bdpm{\begin{displaymath}}
\def\edpm{\end{displaymath}}
\def\nuc#1#2#3 {Nucl. Phys. {\bf#1}, #2 (#3)}
\def\mpla#1#2#3 {Mod. Phys. Lett. A {\bf#1}, #2 (#3)}
\def\plb#1#2#3 {Phys. Lett. B {\bf#1}, #2 (#3)}
\def\prd#1#2#3 {Phys. Rev. D {\bf#1}, #2 (#3)}
\def\prl#1#2#3 {Phys. Rev. Lett. {\bf#1}, #2 (#3)}
\def\ptp#1#2#3 {Prog. Theor. Phys. {\bf#1}, #2 (#3)}
\def\rmp#1#2#3 {Rev. Mod. Phys. {\bf#1}, #2 (#3)}
\def\zpc#1#2#3 {Z. Phys. C {\bf#1}, #2 (#3)}
\def\ibid#1#2#3 {{\it ibid.} {\bf#1}, #2 (#3)}
\def\none#1#2#3 {{\bf#1}, #2 (#3)}
\newcommand{\B}{\textit{B}}
\newcommand{\D}{\textit{D}}
\newcommand{\K}{\textit{K}}
\newcommand{\BABAR}{\textit{BABAR}}
\newcommand{\QED}{\textbf{QED}}

\newcommand{\HQET}{\textbf{HQET}}
\newcommand{\VMDM}{\textbf{VMDM}}
\newcommand{\rJ}{r_{J/\psi}}
\newcommand{\rK}{r_K}
\newcommand{\re}{r_{\gamma}}
\newcommand{\rr}{r_{\rho}}
\newcommand{\rD}{r_D}

\newcommand{\et}{{\it et al.}}

\begin{document}
\begin{titlepage}
\begin{flushright}
UMN-TH-2330/04 \\
November 2004
\end{flushright}
\vspace{0.7in}
\begin{center}
{\Large \bf QED corrections to isospin-related decay rates of charged and neutral B mesons \\
} \vspace{1.2in} {\bf Soo-hyeon Nam\\ }
School of Physics and Astronomy, University of Minnesota, \\
Minneapolis, MN 55455, USA \\
\vspace{1.2in}
\end{center}

\begin{abstract}
We estimate the isospin-violating \QED\ radiative corrections to the charged-to-neutral ratios
of the decay rates for $B^+$ and $B^0$ in non-leptonic $B$ meson decays.
In particular, these corrections are potentially important for precision measurement of
the charged-to-neutral production ratio of \B\ meson in $e^+e^-$ annihilation.
We calculate explicitly the \QED\ corrections to the ratios of two different types of decay rates
$\Gamma(B^+ \rightarrow J/\psi K^+)/\Gamma(B^0 \rightarrow J/\psi K^0)$ and
$\Gamma(B^+ \rightarrow D^+_S \bar{D^0})/\Gamma(B^0 \rightarrow D^+_S D^-)$
taking into account the form factors of the mesons based on the vector meson dominance model,
and compare them with the results obtained for the point-like mesons.
\end{abstract}

\end{titlepage}

 At present \B-factory experiments, a measurement of the $B^+B^-$ to $B^0\bar{B^0}$ production ratio
in $e^+e^-$ annihilation is essential for accurate determination of standard model parameters
such as quark-mixing matrix elements.  The \B\ mesons are produced in pairs at the $\Upsilon(4S)$ resonance,
and the charged-to-neutral production ratio given by
\beq
R^{+/0} = 1 + \delta R^{+/0} = \frac{\Gamma\left(\Upsilon(4S) \rightarrow B^+B^-\right)}
  {\Gamma\left(\Upsilon(4S) \rightarrow B^0\bar{B^0}\right)}
\eeq
has been recently measured by CLEO \cite{CLEO}, \BABAR\ \cite{BABAR},
and Belle \cite{Belle}.  The experimental values of $R^{+/0}$ typically range from 1.01 to 1.10
and the latest significant measurement gives $R^{+/0}_{expr} = 1.006\pm 0.036\pm 0.031$ \cite{BABAR}.\footnote{
It can also be mentioned that \BABAR\ recently reported their preliminary result of the direct measurement of
the branching fraction $\mathcal{B}(e^+e^- \rightarrow B^0\bar{B^0})=0.486\pm 0.010 \pm 0.009$
using partial reconstruction of the decay $\bar{B^0}\rightarrow D^{*+}l^-\bar{\mu}_l$ \cite{BABAR2}.}
The theoretical prediction of the deviation $\delta R^{+/0}$ has been first made by calculating the Coulomb
interaction between point-like mesons \cite{Atwood}, and reexamined later by considering the effects of
the electromagnetic form factors of the mesons and their interaction vertex \cite{Lepage}.
The theoretical values of the ratio $R^{+/0}$ estimated so far range from 1.03 to 1.25.
As well as the Coulomb interactions, it was also shown that the strong interaction phase in the region of a strong resonance
can modify the experimental result of the ratio $R^{+/0}$ near threshold in $e^+e^-$ annihilation \cite{Voloshin04}.

  The practical measurements of the ratio $R^{+/0}$ use the yields of events
for the exclusive processes where the ratio of the decay rates for $B^+$ and $B^0$,
\beq
\frac{\Gamma(B^+ \rightarrow X^+)}{\Gamma(B^0 \rightarrow X^0)} \equiv 1 + \delta^{+/0}(X),
\eeq
is determined by the isotopic invariance. An example of such isospin-related decays is provided by
$B^+ \rightarrow J/\psi K^+$ and $B^0 \rightarrow J/\psi K^0$ where the deviation $\delta^{+/0}(J/\psi K)$
due to the isospin-violating mass differences between these two decays is about 0.0014.
Since this isospin violation is very small and doesn't exceed the present experimental accuracy,
it has been assumed that $\Gamma(B^+ \rightarrow J/\psi K^+)=\Gamma(B^0 \rightarrow J/\psi K^0)$.
In decays of this type, however, there is still an isospin violation due to \QED\ radiative corrections.
At a non-zero deviation $\delta^{+/0}$, the actual value of the production ratio
should be corrected to $R^{+/0} = R^{+/0}_{expr}/(1+\delta^{+/0})$.
For precision measurement in present and future experiments,
it is important to understand the production ratio $R^{+/0}$ of specific exclusive modes
with accuracy better than 1\%.  Especially, the decay mode such as $B^0 \rightarrow J/\psi K^0$ requires
a better understanding due to its great importance in the measurement of $CP$ violation.
Hence, the deviation $\delta^{+/0}(X)$ for a specific decay mode needs to be evaluated more carefully.

 The isospin violation in $B \rightarrow D$ semileptonic decays was discussed earlier
in Ref. \cite{Voloshin98}.  In the decays $B^- \rightarrow D^0 l\tilde{\nu}$ and
$B^0 \rightarrow D^+ l\tilde{\nu}$, for example, the kinematic difference due to the isotopic mass splitting of
the \D\ mesons in the spectra is not negligible
and can amount to a sizeable fraction of 1\%, while the difference is compensated in the total decay rate
with a much better accuracy as predicted by the heavy quark effective theory (\HQET).
On the other hand, in the nonleptonic decays where isospin violation due to the mass difference
between the charged and neutral mesons is expected to be very small,
the \QED\ corrections may not be negligible and dominantly lead to the ratio of
the corresponding decay rates being off unity.  In the standard model, the \QED\ corrections usually amount
to an $\alpha$ order where $\alpha$ is the fine-structure constant and differ by sizeable amounts
depending on the final state masses.
In this paper, we mainly concentrate on the two different types of the  non-leptonic \B\ decays,
$B^+ \rightarrow J/\psi K^+$ and $B^0 \rightarrow J/\psi K^0$ ($VP$ type)
decays as well as $B^+ \rightarrow D^+_S \bar{D^0}$ and $B^0 \rightarrow D^+_S D^-$ ($PP$ type) decays
where $P(V)$ denotes a pseudoscalar(vector) meson, and evaluate explicitly the \QED\ radiative corrections
to the ratios of those decay rates.

\begin{figure}[!hbt]
\centering%
    \includegraphics[width=3cm]{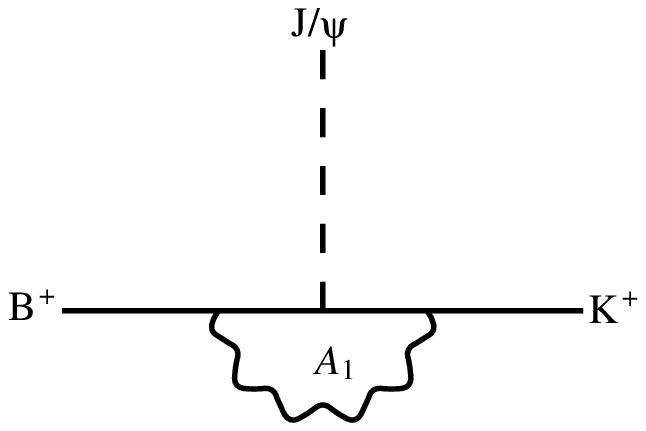} \
    \includegraphics[width=3cm]{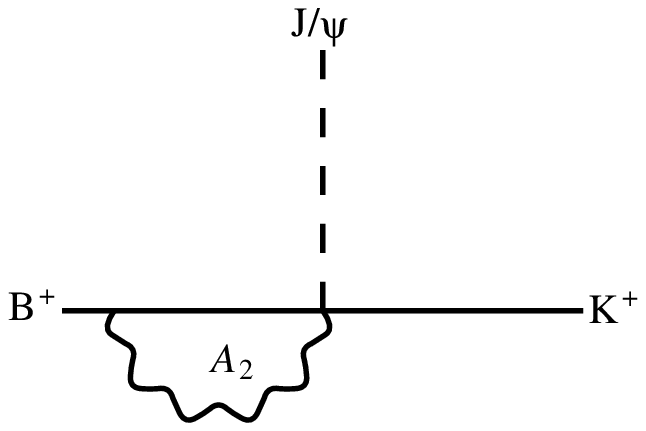} \
    \includegraphics[width=3cm]{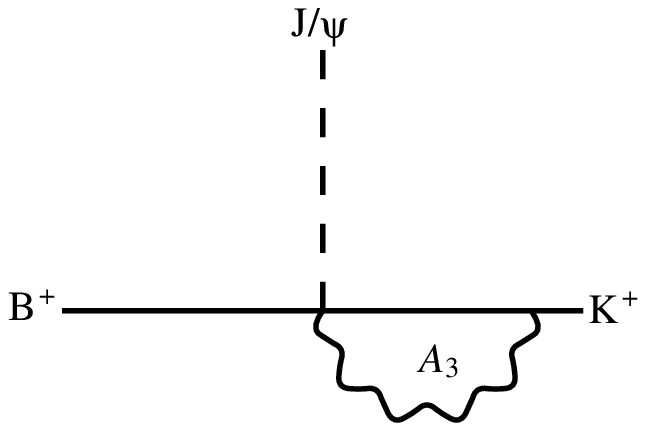} \
    \includegraphics[width=3cm]{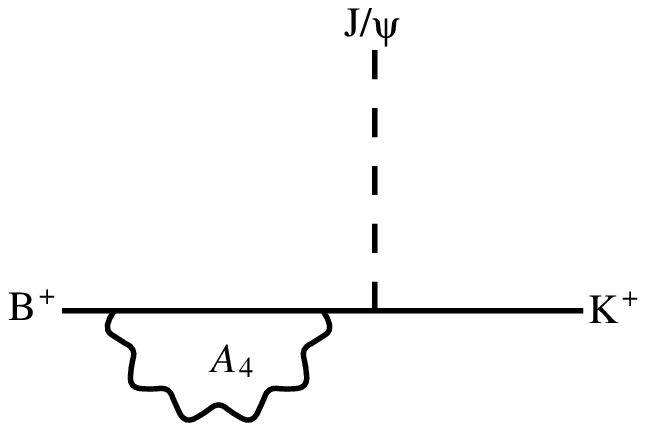} \
    \includegraphics[width=3cm]{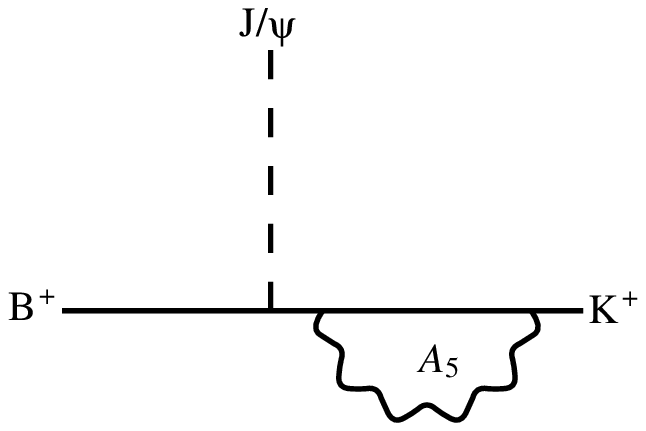}
\caption{One-loop diagrams for $B^+ \rightarrow J/\psi K^+$ decays.}
\label{rad1}
\end{figure}

\begin{figure}[!hbt]
\centering%
    \includegraphics[width=3cm]{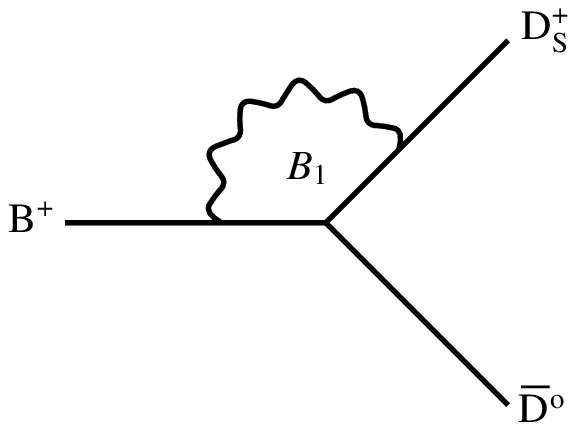} \quad
    \includegraphics[width=3cm]{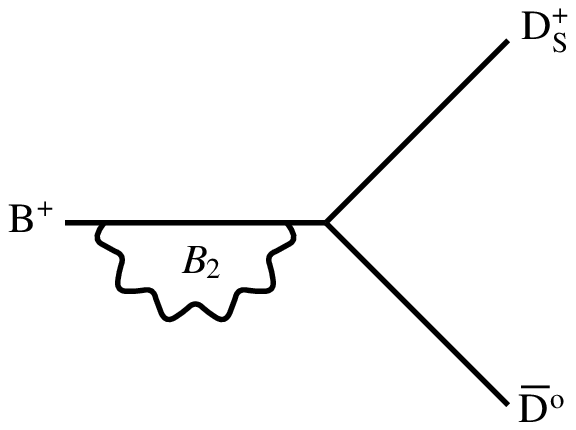} \quad
    \includegraphics[width=3cm]{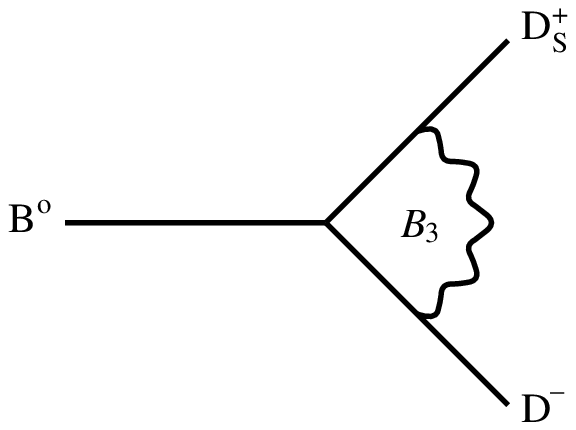} \quad
    \includegraphics[width=3cm]{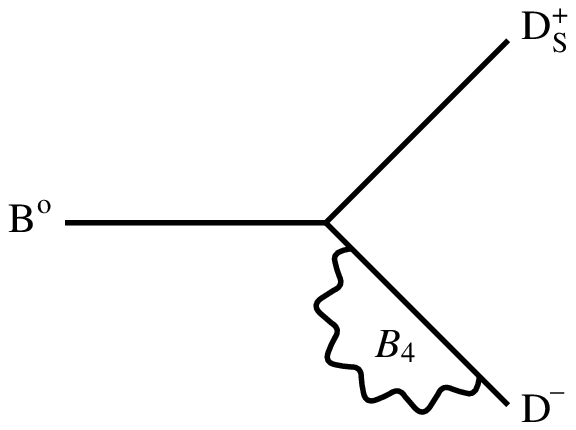}
\caption{One-loop diagrams for $B^+ \rightarrow D^+_S \bar{D^0}$ and $B^0 \rightarrow D^+_S D^-$ decays.
The $D^+_S$ self-energy makes no contribution to the ratio of the corresponding decay rates.}
\label{rad2}
\end{figure}

 The calculation of the \QED\ radiative corrections due to virtual- and real-photon in the leading order
for point-like charged scalar particles is quite straightforward.  The radiative corrections
due to virtual-photons can be obtained by performing the loop integrals shown
in Fig. \ref{rad1} and Fig. \ref{rad2}.
However, one cannot simply treat the mesons as point-like particles
because the \QED\ corrections are sensitive to the structure of the mesons, and thus we require
models for the form factors of mesons.  The electromagnetic form factor $F_X$ of a meson $X$ is successfully
modeled in the intermediate energy region using the vector meson dominance model (\VMDM) \cite{Sakurai}.
The most commonly used \VMDM\ assumes that all photon-hadron coupling is mediated by vector mesons
and the form factors are expressed through their propagators.
Taking account of the electric charges of the quarks comprising the mesons,
we use the \VMDM\ based parametrization for the form factors in such a way that each meson behaves
like a point-like particle in a zero photon momentum limit as follows:\footnote{
We assume here that the form factors are primarily associated with the $\rho$ and $\omega$ mesons,
and neglect the $\phi$ meson contributions for simplicity, which does not change our result significantly.
In other words, we take into account the form factors for the electromagnetic current
of the $u$ and $d$ quarks $(2/3)\bar{u}\gamma_\mu u - (1/3)\bar{d}\gamma_\mu d$, and treat the current
of the $s$, $c$, and $b$ quarks as having a point-like form factor.}
\beqr
F_{B^+}(k^2) &\simeq&  \frac{1}{3} + \frac{2}{3}\frac{m_\rho^2}{m_\rho^2-k^2} \cr
             &\simeq& F_{K^+}(k^2)\ \simeq\ F_{B^+ K^+ J/\psi}(k^2) , \cr
F_{B^0}(k^2) &\simeq& \frac{1}{3} - \frac{1}{3}\frac{m_\rho^2}{m_\rho^2-k^2} \cr
             &\simeq& F_{K^0}(k^2)\ \simeq\ F_{B^0 K^0 J/\psi}(k^2) , \\
F_{D^-}(k^2) &\simeq& -\frac{2}{3} - \frac{1}{3}\frac{m_\rho^2}{m_\rho^2-k^2} , \cr
F_{\bar{D^0}}(k^2) &\simeq&  -\frac{2}{3} + \frac{2}{3}\frac{m_\rho^2}{m_\rho^2-k^2}, \cr
F_{D_S^+}(k^2) &\simeq& 1 , \nonumber
\eeqr
where $k$ is the momentum of photon, and $m_\rho$ is the mass of $\rho$ meson.
Then one can obtain the leading virtual-photon corrections to the ratios of the decay rates
by replacing the photon propagator $1/k^2$ with the modified propagators as follows:
\beqr
\int_{A_i, B_{1,2}}\frac{d^4k}{k^2} &\rightarrow&
    \int_{A_i, B_{1,2}} d^4k \left( \frac{1}{k^2} - \frac{1}{k^2 - m_\rho^2}
    + \frac{m_\rho^2}{3}\frac{\partial}{\partial m_\rho^2}\frac{1}{k^2 - m_\rho^2} \right) , \cr
- \int_{B_3}\frac{d^4k}{k^2} &\rightarrow&
    - \int_{B_3} d^4k \left( \frac{1}{k^2} - \frac{1}{k^2 - m_\rho^2} \right) , \\
- \int_{B_4}\frac{d^4k}{k^2} &\rightarrow&
   - \int_{B_4} d^4k \left( \frac{1}{k^2} - \frac{1}{k^2 - m_\rho^2}
    - \frac{m_\rho^2}{3}\frac{\partial}{\partial m_\rho^2}\frac{1}{k^2 - m_\rho^2} \right) , \nonumber
\label{propagator}
\eeqr
where the subscripts $A_i$ and $B_j$ denote each loop
shown in Fig. \ref{rad1} and Fig. \ref{rad2}.
Note that it is clear from the above equations that UV divergences arising from the one-loop calculation
are naturally eliminated.  For actual calculation, it is convenient to introduce a small non-zero photon
mass $m_\gamma$ and obtain the analytic expressions of the virtual-photon corrections
for the point-like mesons first, so that the additional terms due to the $\rho$ meson propagator
in the \VMDM\ can be simply obtained by replacing the photon mass with the $\rho$ meson mass.
Then, the IR divergences arising from the one-loop diagrams are expressed by $\ln m_\gamma$,
and the linear terms of $m_\gamma$ can be dropped whereas those of $m_\rho$ cannot
because the $\rho$ meson is heavier than the \K\ meson.

 After a straightforward but somewhat lengthy calculation,
the leading virtual-photon corrections to the ratios of the decay rates are given by
\beqr \label{virtual}
\delta_{virt}^{+/0}(J/\psi K) &=& \left\{ \begin{array}{ll}
  \frac{\alpha}{2\pi}\big(f_K(\rJ, \rK; \re) + O(\rK)\big)
   & \mbox{\ \ (point-like mesons)}\\
  \frac{\alpha}{2\pi}\big(g_K(\rJ, \rK, \rr; \re) + O(\rr)\big)
   & \mbox{\ \ (\VMDM)} \end{array} \right. , \\
\delta_{virt}^{+/0}(D^+_SD) &=& \left\{ \begin{array}{ll}
  \frac{\alpha}{2\pi}f_D(\sqrt{1-4\rD}; \re)
   & \mbox{\ \ (point-like mesons)}\\
  \frac{\alpha}{2\pi}\big(g_D(\sqrt{1-4\rD}, \rr; \re) + O(\rr)\big)
   & \mbox{\ \ (\VMDM)} \end{array} \right. ,
\eeqr
where
\beqr
f_K(x,y;w) &=&-4+2\big(2-\ln(1-x)\big)\ln(1-x) - 2\textrm{Li}_2(x) \cr
           &&-\left(1-\frac{1}{2}\ln y\right)\ln y +\left(\ln\frac{(1-x)^2}{y}-2\right)\ln w , \label{f_K}\\
g_K(x,y,v;w) &=& -\frac{2}{3}+ \frac{1}{3}\ln\frac{(1-x)^2}{y}
           +\frac{5\pi}{4}\sqrt{\frac{v}{y}}\left(1-\frac{(3+x)\sqrt{y}}{1-x}+\frac{v}{5y}\right) \cr
          &&+\frac{v}{12y}\left[1-\frac{3v}{y}-\left(10-\frac{v}{y}\right)\ln\frac{v}{y}\right]
          +\left(\ln\frac{(1-x)^2}{y}-2\right)\ln\frac{w}{v} , \label{g_K}\\
f_D(x;w) &=& \frac{1+x^2}{1-x^2}\ln\frac{4}{1-x^2}+\frac{\pi^2(1+x^2)}{3x}
          -\frac{2(2+x^2)}{x}\textrm{Li}_2\left(\frac{1-x}{1+x}\right) \cr
          &&+\frac{2}{x}\textrm{Li}_2\left(\left(\frac{1-x}{1+x}\right)^2\right)
          -\frac{2}{x}\ln\frac{1+x}{1-x}\ln\frac{2}{1+x}  \label{f_D}\\
          &&+x\ln\frac{1+x}{1-x}\left(2\ln x -\frac{3-x^2}{1-x^2}-\frac{1}{2}\ln\frac{1+x}{1-x} - \ln w \right) , \cr
g_D(x,v;w) &=& -\frac{2}{3}+ \frac{1}{3x}\ln\frac{1+x}{1-x}
         -\frac{\pi\sqrt{v}}{6x^2}\left[5\left(1-\frac{x^2}{2}\right)+\frac{7(1+2x^2)}{\sqrt{1-x^2}}\right] \label{g_D}\\
          &&+\frac{v}{3(1-x^2)}\left(17+4\ln\frac{4v}{1-x^2}\right)-x\ln\frac{1+x}{1-x}\ln\frac{w}{v} , \nonumber
\eeqr
and where $r_P\equiv m_P^2/m_B^2$ and we use $m_{P^+}=m_{P^0}\equiv m_{P}$.
In the above equations, $\ln r_K$ term in Eq. (\ref{f_K}) has a mass singularity
in the limit $m_K \rightarrow 0$ and it is removed by the same term in the real-photon corrections.
In the \VMDM\ there are additional terms such as $m_\rho/m_K$ for non-zero $m_\rho$ and $m_K$ in general,
and Eq. (\ref{g_K}) was obtained for the \K\ meson satisfying $m_\rho/2 < m_K < m_\rho$.
In the limit $m_K \rightarrow 0$, the additional singular terms $\ln m_K$ and $m_\rho/m_K$ in Eq. (\ref{g_K})
produced by the $\rho$ meson propagator are replaced by $\ln r_\rho$,
and the function $g_K$ can be rewritten as
\beqr
g_K(x,y,v;w) &=& -\frac{3}{4}+\frac{5\pi^2}{6}+2\left(\frac{1}{3}-\ln x\right)\ln(1-x)-2\textrm{Li}_2(x)
                 -2\textrm{Li}_2(1-x)  \cr
           &&-\left(1-\frac{1}{2}\ln y\right)\ln y+\left(\frac{8}{3}-2\ln(1-x)+\frac{1}{2}\ln v\right)\ln v  \\
           &&-\frac{5\pi(3+x)}{4(1-x)}\sqrt{v}+\left(\ln\frac{(1-x)^2}{y}-2\right)\ln w . \nonumber
\eeqr
As for the $D$ mesons, one can see the term $m_\rho/m_D$ in Eq. (\ref{g_D}) and this formula
is valid for any other scalar meson $P$ satisfying the condition $m_P > m_\rho$.

The real-photon corrections can be simply obtained by assuming that all mesons behave like point-like particles.
\beqr  \label{real}
\delta_{real}^{+/0}(J/\psi K) &=& \frac{\alpha}{2\pi}\big(h_K(\rJ, \rK; \re) + O(\rK)\big) , \cr
\delta_{real}^{+/0}(D^+_SD) &=& \frac{\alpha}{2\pi}h_D(\sqrt{1-4\rD}; \re) ,
\eeqr
where
\beqr
h_K(x,y;w) &=& 9-\frac{\pi^2}{3}-4\sqrt{x}-\frac{2(1-3x)}{(1-x)^2}-2\big(3-\ln (1-x)\big)\ln(1-x)  \cr
            &&-\frac{2x(1-4x+x^2)\ln x}{(1-x)^3}+\left(1-\frac{1}{2}\ln y\right)\ln y \\
            &&-\left(\ln\frac{(1-x)^2}{y}-2\right)\ln w  , \cr
h_D(x;w) &=& 2x\left[\frac{\pi^2}{3}-\textrm{Li}_2\left(\frac{1-x}{1+x}\right)
             -\textrm{Li}_2\left(\left(\frac{1-x}{1+x}\right)^2\right)\right] \cr
           &&+x\ln\frac{1+x}{1-x}\left(1+2\ln x + 6\ln\frac{2}{1+x}+\frac{1}{2}\ln\frac{1+x}{1-x}+\ln w\right) .
\eeqr
Now one can obtain the total \QED\ radiative corrections $\delta_{\QED}^{+/0}=\delta^{+/0}_{virt}+\delta^{+/0}_{real}$,
and explicitly see that the IR divergent terms expressed by $\ln r_\gamma$
in the virtual-photon corrections cancel out the same terms arising in the real-photon corrections
in a usual way.

Using the standard values of the meson masses, we estimate the numerical values of the \QED\ radiative corrections to
the ratios of the decay rates:
\beqr
\delta_{\QED}^{+/0}(J/\psi K) &\simeq& \left\{ \begin{array}{ll}
   -0.0012 & \mbox{\ \ (point-like mesons)}\\
   0.0014 & \mbox{\ \ (\VMDM)} \end{array} \right. , \label{num1}\\
\delta_{\QED}^{+/0}(D^+_SD) &\simeq& \left\{ \begin{array}{ll}
   0.0115 & \mbox{\ \ (point-like mesons)}\\
   -0.0013 & \mbox{\ \ (\VMDM)} \end{array} \right. \label{num2}.
\eeqr
As a comparison, we list the radiative corrections for the point-like mesons too.
In $B\rightarrow J/\psi K$ decays, the \QED\ radiative correction to the ratio of the decay rates is small
because the real-photon correction is offset by the virtual-photon correction.
On the other hand, the \QED\ correction in $B\rightarrow D_S^+ D$ decays is sizable for the point-like mesons
but significantly reduced in the \VMDM\ due to the similar compensation between the real- and virtual-photon
correction.  In both cases, the deviations $\delta^{+/0}$ due to the isotopic mass splitting
of the $K$ and \D\ mesons are about 0.0014, so that the total isospin violation effect
in $B\rightarrow J/\psi K$ decays becomes about 0.3\% in the \VMDM\ while that in $B\rightarrow D^+_S D$ decays
becomes negligible. In different decay modes, of course, the corrections can be enhanced or reduced
due to different final state masses.
Therefore, in order to obtain the production ratio $R^{+/0}$ more accurately
in precision measurements, one should include the correction $\delta^{+/0}$
or find an appropriate decay mode where the isotopic violation due to the \QED\ effects is very small.
As well as the determination of $R^{+/0}$, the obtained results are also of importance in the study of the structure
of hadron.  We explicitly show in Eq. (\ref{num1}) and Eq. (\ref{num2}) how the form factors of the mesons
affect the radiative corrections.  This work can be used to test the model itself and will be useful
for the study of the form factors when future experiments are available to access the ratios of the decay rates.

 The author would like to thank M.B. Voloshin for suggesting this problem and for helpful discussions,
and S. Rudaz for his useful comments.
 This work is supported in part by the DOE grant DE-FG02-94ER40823.

\newpage

\end{document}